\documentstyle[11pt,twoside,newpasp,epsf,natbib]{article}
\def\edcomment#1{\iffalse\marginpar{\raggedright\sl#1\/}\else\relax\fi}
\reversemarginpar

\newcommand{\mem}[1]{\mathrm{ #1}}

\newcommand{\kap}[1]{Sect.\,\ref{#1}}
\newcommand{\msun}{\, {\rm M}_\odot}
\newcommand{\czw}{^{12}\mem{C}}
\newcommand{\cdr}{^{13}\mem{C}}

\newcommand{\ose}{^{16}\mem{O}}

\newcommand{\spr}{\mbox{$s$-process}}
\newcommand{\zrse}{^{96}\mem{Zr}}
\newcommand{\zrvi}{^{94}\mem{Zr}}

\begin{document}
\title{Current models for the evolution of AGB stars}
 \author{Falk Herwig}
\affil{University of Victoria, Victoria, BC, Canada}

\begin{abstract}
While the basic properties of AGB stellar evolution are well
established, comprehensive observational studies of late phases of
intermediate mass stars continue to generate puzzles for
current stellar models. Here, I review current techniques to model
AGB stars, and I discuss important aspects of current research of AGB
(and post-AGB) stellar  evolution with a particular focus on how these
interrelate.
\end{abstract}

\section{Introduction}
\label{sec:intro}

The final evolution of low and intermediate mass stars proceeds
through the Asymptotic Giant Branch (AGB), post-AGB phase and finally
the white dwarf phase. The post-AGB phase can be sub-divided according
to the evolutionary state of the circumstellar matter into the central
star phase of proto-planetary nebulae and planetary nebulae (PNe). 

By the early eighties of the last century the dominant aspects of the
evolution of low and intermediate mass giant stars had been qualitatively
recognized and summarized in the seminal review by Iben \& Renzini (1983).
These include the first and second dredge-up (DUP),
thermal pulses (TP), the relation of core mass, luminosity, metalicity
etc., the luminosity variation during a pulse cycle, the main
nuclear reactions and element production, hot bottom burning (HBB,
called more prosaic \emph{envelope burning} at that time), pulsational
properties, mass loss (including the necessity for some kind of
\emph{superwind}), the concept of synthetic AGB models, the AGB
luminosity function as well as the large potential of pre-solar
meteoritic material which originates in the cool outflows of AGB
stars. In addition the general picture of the transition from the AGB
to the post-AGB, the formation of PNe was
established and so were the concept of post-AGB TPs and the
nature of the PNe abundances as reflecting the heritage of the progenitor
evolution. It was speculated on the role of overshooting from the
bottom of the convective envelope and the need for rotationally induced
mixing. And already at that time IRC 10+216 had received much attention.

Extensive grids of full stellar evolution models (see
\kap{sec:mod}) including the above mechanism in more detail 
were constructed over the following  
decade (Lattanzio, 1986; Boothroyd \& Sackmann, 1988; 
Vassiliadis \& Wood, 1993; Bl\"ocker, 1995, 
and related papers by these
authors). 
AGB stars  were recognized as the production site 
for the \spr\ elements, but the dominance of the radiative burning of
$\cdr$ during the interpulse phase was only discovered later
(Straniero et\,al., 1995; Gallino et\,al., 1998). In addition a deeper understanding
of the third DUP in low mass AGB stars as required by the
luminosities of carbon stars in the LMC (Richer, 1981) developed
only fairly recently when the dependence on numerical details
(Frost \& Lattanzio, 1996) and the connection to the treatment of the
convective boundaries of \emph{both the bottom of the convective
  envelope and the bottom of the He-flash convection zone} was recognized
(Herwig et\,al., 1997; Mowlavi, 1999; Herwig, 2000).  
The more recent efforts in modeling AGB stars were discussed by
several authors during the IAU Symp.~191 in Montpellier, France in
1998 (e.g.\ Bl\"ocker, 1999).

It turns out that we now know much more about the details of AGB
stellar evolution, in particular with respect to the nucleosynthesis
in these stars
(e.g.\ Forestini \& Charbonnel, 1997; Mowlavi \& Meynet, 2000; 
Busso et\,al., 1999; Marigo, 2001).
Degenerate TPs were described by Frost et\,al.\ (1998), and
extensive computations have explored the evolution of the 
most massive
AGB stars in the initial mass range of 9 to 11$\msun$
(Ritossa et\, al., 1999, and references therein). The swallowing of
planets or brown dwarfs by AGB stars have been investigated by 
Siess \& Livio (1999).

In this paper I will not repeat the basic properties of AGB stars,
which are - as mentioned above - well documented in the
literature. Instead, I describe in \kap{sec:mod} the different methods
currently in use for modeling the interior processes of AGB stars.
In \kap{sec:connection} I concentrate on the \emph{various links between the
different aspects of the advanced evolution} of low and intermediate
mass stars and how 
these relations offer new diagnostic tools for investigating AGB stars. 

\section{Modeling AGB stars}
\label{sec:mod}

\paragraph{Full stellar models} In this approach the  basic stellar
structure equations are solved  through the entire star for each time step. The
change of abundances due to both mixing and nuclear processing is
computed at each time step as well. Mixing is generally treated in
some time dependent manner (e.g.\ by solving a diffusion-like equation
for each chemical species), while nucleosynthesis is treated by 
solving the set of rate equations which describe nuclear production
and destruction for each isotope. The evolution is  followed
consistently from the pre-main sequence through all evolutionary
phases up to the TP-AGB, and sometimes continued into the white dwarf
stage. Models have high spatial and time resolution. A thermal pulse cycle
is resolved by of the order of 5000 models on about 2000 spatial grid
points. In particular high resolution is needed in time during the DUP
phase ($\Delta t \sim$ weeks) and in space at the core-envelope
interface ($\Delta m \sim 10^{-6}\msun$). Recent computations of such
models include those by Straniero et\,al.\ (1997), Wagenhuber \& Groenewegen (1998) and Heriwg (2000).

 While most models have been
computed using the operator split method in which the treatment of
structure, mixing and nucleosynthesis is separately and sequentially
treated, there have been efforts to overcome this
approximation. Pols \& Tout (2002) have computed   TP-AGB models 
with an updated version of the Eggleton code which solves the
structure equations and those of composition changes
simultaneously. Their results are good and bad news. The good news is
that the model properties with this much more computationally
demanding strategy is very similar to recent models by
Mowlavi (1999) and Herwig (2000) in particular with respect
to the amount of third DUP obtained in the models. The bad news is that
results are still somewhat dependent on the detailed treatment of
mixing at the convective boundary (see the paragraph on the third
DUP in \kap{sec:connection}). 

Two other variations on the operator split theme have been
played. Straniero et\,al.\ (1997)  have combined  the structure and the 
nuclear burning operator but the implications for the evolution of AGB
stars are not immediately clear. More relevant for AGB and post-AGB
stars is the combination of the mixing and the nucleosynthesis
operator. Such a treatment is needed whenever a relevant nuclear time
scale is similar or smaller than the convective turnover
timescale, which applies to models of the lithium evolution in HBB
stars, zero-metalicity AGB stars as well as the \emph{very late}
variant of the post-AGB TP.

Models as described in this category are now outfitted with a
description of the effects of stellar rotation. The pilot study by
Langer et\,al.\ (1999) showed considerable effect both on the structural
and the abundance evolution. In particular strong shear mixing induced
by a steep angular velocity gradient at the
core-envelope interface after the onset of the third DUP is
potentially instrumental for the $\cdr$ neutron source of the \spr.

\paragraph{Synthetic models} 
The detailed computation of full TP-AGB stellar models has been for a long
time too tedious a task to compute complete sets of models for a wide
range of mass, metalicity, mass loss and extra mixing prescription,
with a sufficiently small spacing in all these parameters and with
many isotopic species needed for detailed comparisons 
with observations. An alternative are synthetic models, which summarize
the results of full stellar models through simple analytical
relations. This procedure has been initially applied by
Iben \& Truran (1978) and Renzini \& Voli (1981). 
Such models are useful for other than the AGB evolutionary
stages as well (Hurley et\,al., 2000). Modern synthetic models
with a focus on AGB stars have been presented by
Groenewegen \& de Jong (1993)  and Marigo et\,al.\ (1996) and subsequent
improvements. Most recent models feature a full 
envelope integration which
allows for a mass and metalicity dependent parameterization of the
third dredge-up and for a reliable treatment of the HBB burning in
massive AGB stars (Marigo, 1998; Mariga et\,al., 1999). 
Because it is possible
to cover a large parameter space, extensive tables of
stellar yields have been computed 
(van den Hoek \& Groenewegen, 1997; Marigo, 2001).
 These tables also include 
predictions on the PN composition. In principle, it should be simple
to consider the effect of an oxygen enriched intershell, as studied in
detail by Herwig (2000).

\paragraph{Post-processing and parametric nucleosynthesis models}
Most modern stellar evolution codes consider a sufficient
number of isotopes in a nuclear network in order to follow the main
nuclear processes in some detail. For AGB stellar models these include
hydrogen-burning by the CNO cycle, and triple-$\alpha$ and
$\czw(\alpha,\gamma)\ose$ for He-burning. However, AGB stars are
interesting because of their large potential of interesting
nucleosynthesis. Because of the time consuming nature of computing
stellar structure model sequences of AGB stars, the method of
post-processing can be employed in order to create detailed
nucleosynthesis models of AGB stars with large networks.
A post-processing code uses structure output of full stellar evolution
models (see above). For a given initial composition
(typically the abundance profile of the first post-processed structure
model) the abundance changes are computed model by model according to
the sequence of stellar structure models. Such a method may be
confined only to a particularly interesting section of a star, like
the layer including the two burning shells and the intershell in AGB
stars. The post-processing method is, for example, routinely used in
the combination of the  Australian \emph{MSSSP/MOSN} codes
(Lattanzio et\,al., 1996). Gallino et\,al. (1998) and 
Goriely \& Mowlavi (2000) presented
post-processing models for the 
\spr\ in AGB stars. Here this method is useful because the decisive 
nucleosynthesis happens only in a tiny mass layer in the star, to which
modeling can be confined. In addition small artificial changes can be
introduced, which help to overcome certain incapabilities of mixing
in full models.
Finally, Herwig et\,al. (2002) used this method in a high
resolution variant to study the \spr\ in rotating AGB stars. Since
mixing plays an important role in the considered nuclear production
site, their fully implicit code also solves for mixing processes
according to the input stellar structure models.

\section{Making the connection: Concepts across the final stages of evolution}
\label{sec:connection}

Several concepts and properties of AGB and
post-AGB stars and their relations are displayed in Fig.~1. In this 
\begin{figure}
  \plotone{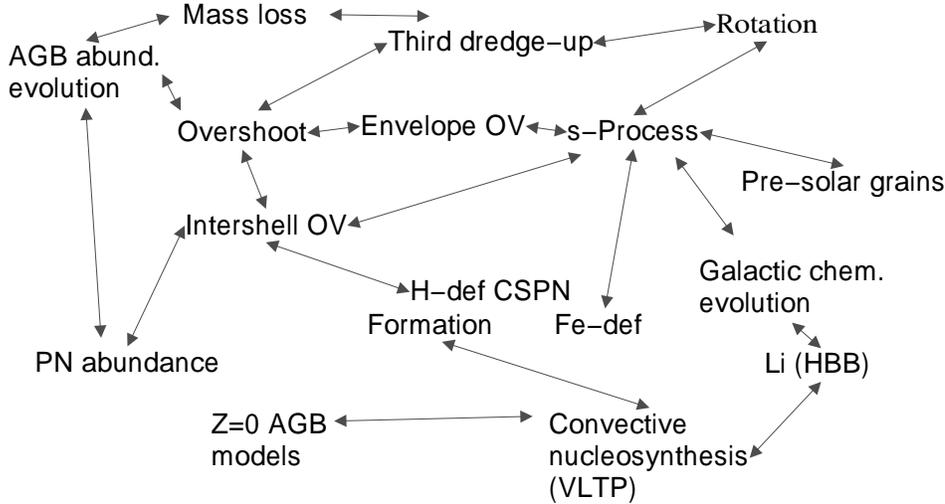} \vspace{-7em}
\caption{The mind map of advanced
  evolutionary phases including AGB stars shows associations between
  their properties and concepts. \textsf{OV} means overshoot.}
\end{figure}
section we will discuss the displayed aspects.

How to obtain \textbf{the third DUP} in TP-AGB models has been
discussed intensely in recent years.  Many models still do not show
DUP at low core masses (e.g.\ Wagenhuber \& Groenewegen, 1998).
The models by Straniero etal. (1997) 
feature DUP by using no extra mixing, but they
generate \textbf{carbon stars} too luminous to be compatible with the
C-star luminosity function (CLF). It is clear now that
\textbf{envelope overshooting} improves the models ability to
reproduce efficient DUP at low core masses (Mowlavi, 1999), and also
the operator split issue plays a role (see \kap{sec:mod}).
Nevertheless, also for these models the DUP is probably not compatible
with the CLF (although this has not been checked quantitatively), and
the same is true for the investigations of the dredge-up law 
by Karakas etal. (these proceedings; Lattanzio, priv.\ com.). DUP can however
be enhanced beyond the level achieved in the above 
mentioned models by \textbf{overshoot from the He-flash
  convection zone} (intershell overshoot, Herwig, 2000). The
set of model calculations presented by Herwig et\,al.\ (2000)
 contain
carbon stars with the lowest luminosities obtained from full stellar
models. However, a quantitative comparison with the CLF has not yet
been carried out. Current models of rotating AGB stars (Lanter et\,al., 1999)
do not show enhanced DUP efficiency.
The influence of rotation on the \spr\ in AGB stars has been addressed
by Herwig, Langer \& Lugaro (these proceedings).

The above mentioned \textbf{intershell overshoot} has two side effects
which appear to contradict each other. Intershell overshoot 
increases the temperature at the bottom of the He-flash convection
zone and thereby leads to $\zrse/\zrvi$-ratios incompatible with
measurements of \textbf{mainstream pre-solar SiC grains}
(Lugaro \& Herwig, 2001). In addition intershell overshoot significantly
increases the \textbf{oxygen abundance} in that layer. It can not be
stressed enough that this 
is indeed a very desirable feature if one is concerned with the
origin of \textbf{H-deficient post-AGB stars} of PG1159 and [WC]-type
which are 
known for some time now to have surface oxygen mass fractions in the
range $5 \dots 15\%$ (e.g.\ Hamann, these proceedings). It has been
shown by Herwig et\,al.\ (1999)  that 
this observed O-abundance can only be reproduced by evolution models of
H-deficient post-AGB stars, if the intershell has been pre-enriched
with oxygen already on the AGB. So, we are faced here with an
interesting \emph{case} in which pre-solar meteoritic grains, the
\spr, mixing, stellar structure, expanding atmosphere models and
spectroscopy are all somehow involved. A systematic study of the implications 
of a significant oxygen abundance in the intershell remains to
be done. It is clear that various aspects of AGB and post-AGB
evolution are closely related. Iron-deficiency has now been
established in a number of H-deficient post-AGB stars (e.g.\ Werner,
these proceedings) and this may be
another example in which the progenitor AGB evolution needs to be
carefully included into the interpretation (Herwig, Lugaro, Werner,
these proceedings).

H-deficient PN central stars can form as a result of a post-AGB
He-flash and in some cases this flash involves \textbf{convective
nucleosynthesis} processes where mixing and burning are closely
coupled. This has to be reflected by the numerical treatment of this
phase (Herwig, 2001). Such a coupled algorithm has of course
applications elsewhere, e.g.\ for the well known Li-production in HBB AGB
models (Mazzitelli et\,al., 1999; Ventura et\,al., 2000; 
Bl\"ocker et\,al., 2000). 
Unfortunately, galactic
chemical evolution models for lithium seem to arrive at antipodal
conclusions with respect to the importance of AGB lithium production
site (Travaglio et\,al., 2001; Romano et\,al., 2001). 
This may very likely be related 
to the different treatment of mass loss, in particular during the
final tip-AGB evolution (Lattanzio, priv.\,com.).

 The \textbf{mass loss} law for AGB stars  is still not sufficiently well
understood, despite intense theoretical and observational
efforts. In particular dynamic wind models which consider in great
detail the properties of dust have resulted in theoretical mass loss
formulae ready to use in stellar models of carbon stars
(Arndt et\,al., 1997). However, if such a mass loss formula is applied to
the AGB evolution sequence of Herwig (2000) when the star has
become a carbon star due to DUP the absolute mass loss rate is at that
time equivalent to a Reimer's law parameter of $\eta = 0.5$, indicating
that many more thermal pulses will occur before the tip of the AGB is
reached. In fact, even after about 20 more thermal pulses of the now
carbon rich stellar model the mass loss does not pick up very much.
Synthetic models (see above) on the other hand have constrained the mass 
loss parameter $\eta$ (together with
the DUP parameter $\lambda$ and the minimum core mass for the onset of
DUP) by matching observations of Galactic disk and MC AGB stars. In
this way van den Hoek \& Groenewegen (1997)  find $\eta=4$.  For a lower mass loss rate
carbon stars become too luminous as the core mass increases
from TP to TP over many thermal pulses. We can not expect these two values for
$\eta$ from these totally different approaches to be identical (e.g.\
because of different temperature dependency of the mass loss laws). The
large difference however is indicative that the confrontation of any
new mass loss law with the carbon star luminosity function of the
clouds within full stellar models with dredge-up is a critical test. 

The \textbf{abundance
  evolution} of AGB models depends critically on mixing processes like
overshoot and rotation. The PN abundances are those of the AGB surface
during the final tip AGB evolution. The usefulness of synthetic AGB
models for predicting abundances has already been emphasized. It
should however be noted that these models only contain those effects
which have been considered in the underlying full stellar models. From
H-deficient models of post-AGB stars it is now known that AGB progenitor
models without intershell overshoot (i.e.\ without intershell oxygen
enhancement) are not compatible with observations. An obvious
inconsistency would be to compare the oxygen abundance in a PN with a
[WC]-type central star to current synthetic AGB models.

Finally, peculiar TPs with ingestion of protons into the underlying
He-intershell have been found in \textbf{zero-metalicity AGB models}
 by Cassisi et\,al.\ (1996) and were
rediscovered by Chieffi et\,al. (2001) in zero-metalicity stars.
These H-flashes were confirmed by Siess et\,al.(2002) and by my own
computations. However, Marigo et\,al.\ (2001) do not mention
these convective episodes of the H-shell. As shown by
Straka \& Tscharnuter (2001) numerical methods 
for convective nucleosynthesis may
have to be applied for  Z=0 stellar models, but the importance of this
effects remains to be quantified.
 
\paragraph{Acknowledgments:} I would like to thank D.A. VandenBerg
for support through his Operating Grant from the Natural Science and
Engineering Research Council of Canada. I have also enjoyed very useful 
discussions with John Lattanzio.

\end{document}